\documentclass[aps,prl,twocolumn,superscriptaddress]{revtex4-2}
\usepackage{amsmath,amssymb}
\usepackage{graphicx}
\usepackage{color}
\usepackage{rotating}
\usepackage{bm}
\usepackage{setspace}
\usepackage{hyperref}
\usepackage{multirow}

\hypersetup{
colorlinks=true,
urlcolor= blue,
citecolor=blue,
linkcolor= blue,
bookmarks=true,
bookmarksopen=false,
}

\begin{document}

%\preprint{}

\title{Electronic tuning of the soft-phonon transport anomaly in Ta$_2$Ni(S$_x$Se$_{1-x}$)$_5$}


\author{Yuan-Shan Zhang}
\email{Yuanshan.Zhang@fkf.mpg.de}
\affiliation{Max Planck Institute for Solid State Research, 70569 Stuttgart, Germany}
\author{Masahiko Isobe}
\affiliation{Max Planck Institute for Solid State Research, 70569 Stuttgart, Germany}
\author{Hidenori Takagi}
\affiliation{Max Planck Institute for Solid State Research, 70569 Stuttgart, Germany}
\affiliation{Institute for Functional Matter and Quantum Technologies, University of Stuttgart, 70569 Stuttgart, Germany}
\affiliation{National Institute for Materials Science, 305-0047 Tsukuba, Japan}
\author{Dennis Huang}
\email{D.Huang@fkf.mpg.de}
\affiliation{Max Planck Institute for Solid State Research, 70569 Stuttgart, Germany}

\date{\today}

\begin{abstract}
Ta$_2$NiSe$_5$ continues to be investigated for its phase transition at $T_\textrm{c}$ = 326 K, where it develops both an electronic gap and a distortion of its Ta/Ni chains. One intriguing feature at $T_\textrm{c}$ seen in thermal transport is the giant anisotropic scattering of phonons moving perpendicular to the chains, which is apparently associated with the softening of a transverse acoustic phonon, but whose microscopic origin and significance demand clarification. By tuning the normal-state band overlap/gap with S substitution, we uncover a close connection between this soft-phonon transport anomaly and underlying electronic instabilities: When Ta$_2$Ni(S$_x$Se$_{1-x}$)$_5$ approaches a band insulator at high $x$, and signatures of the electronic transition are suppressed, the soft-phonon transport anomaly concomitantly vanishes. Our results establish the following picture for the Ta$_2$Ni(S$_x$Se$_{1-x}$)$_5$ family: Near the S end, a sole lattice instability gives rise to a weak structural transition with $T_\textrm{c}$ approaching 130 K. Near the Se end, additional electronic instabilities boost $T_\textrm{c}$ up to 326 K and amplify experimental signatures of the transition. The strong interaction between electrons, holes, and the lattice is manifested as a soft-phonon transport anomaly accompanied by electronic fluctuations, which include excitonic and hybridization-gap fluctuations.
\end{abstract}

%\keywords{}

\maketitle

\begin{figure}
\includegraphics[width=\columnwidth]{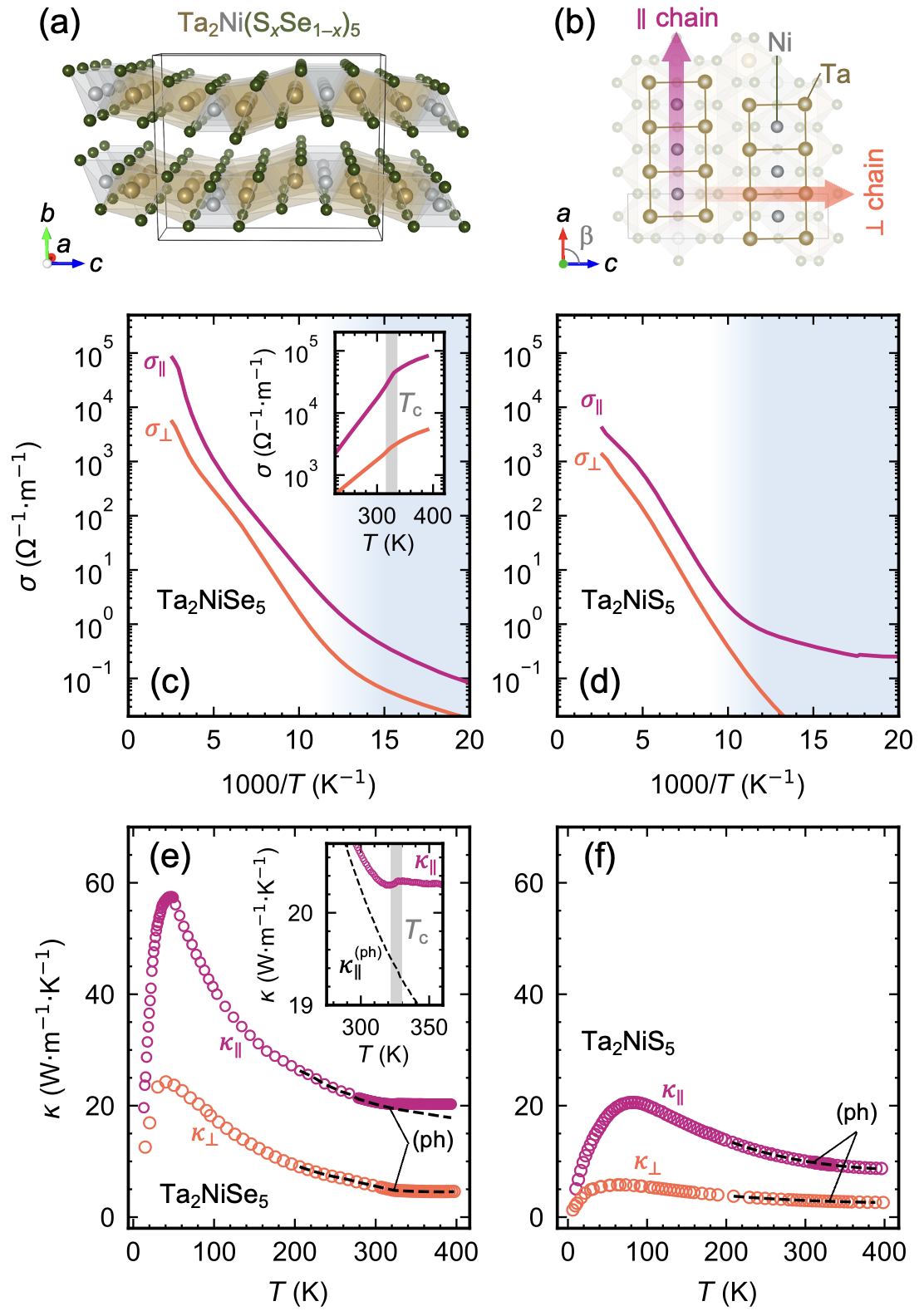}
\caption{(a) Crystal structure of Ta$_2$Ni(S$_x$Se$_{1-x}$)$_5$ with unit cell shown. (b) View along $b$ axis with Ta/Ni chains emphasized. (c), (d) Electrical conductivity $\sigma$ of Ta$_2$NiSe$_5$ and Ta$_2$NiS$_5$ plotted on a logarithmic scale against the inverse temperature $1000/T$. The directions parallel ($\parallel$; along $a$) and perpendicular ($\perp$; along $c$) to the chains are shown in (b). The blue shading marks an extrinsic regime dominated by impurity-band conduction. The inset of (c) plots $\sigma$ vs.~$T$ near the transition temperature $T_\textrm{c}$ of Ta$_2$NiSe$_5$. (e), (f) Thermal conductivity $\kappa$ of Ta$_2$NiSe$_5$ and Ta$_2$NiS$_5$ vs.~$T$. The dashed black curves represent the estimated phonon (ph) contributions. The inset of (e) is an enlargement of $\kappa_{\parallel}$ and $\kappa_{\parallel}^{(\textrm{ph})}$ near $T_\textrm{c}$. The data in (c) and (e) are replotted from Ref.~\cite{Zhang_PRB_2021}.}
\label{Fig1}
\end{figure}

\textit{Introduction}---The quasi-one-dimensional (1D) chalcogenide Ta$_2$NiSe$_5$ [Fig.~\ref{Fig1}(a)] undergoes a phase transition at $T_\textrm{c}$ = 326 K, whose origin remains contested. The transition is both electronic and structural, as evidenced by a simultaneous opening of an electronic gap (0.16--0.30 eV \cite{Lu_NatComm_2017, Lee_PRB_2019, He_2021}) and an orthorhombic-to-monoclinic distortion of its unit cell ($\beta = 90.7^\circ$ \cite{Nakano_PRB_2018}). An early explanation for this transition invoked the concept of an excitonic insulator \cite{Jerome_1967, Wakisaka_2009, Lu_NatComm_2017}. The premise is that the narrow band overlap in Ta$_2$NiSe$_5$ at high temperatures is unstable against the spontaneous formation of excitons, i.e., electron-hole pairs, with the electron and hole likely located on adjacent Ta and Ni sites along a chain [Fig.~\ref{Fig1}(b)], resulting in an insulating ground state with a many-body gap. Here, the concomitant lattice distortion is apparently a secondary effect. A later explanation emphasized the symmetry reduction from orthorhombic to monoclinic point groups due to the minute lattice distortion, which permits the hybridization of Ni 3$d$ valence and Ta 5$d$ conduction bands of formerly opposite parities \cite{Mazza_2020, Subedi_2020, Watson_PRR_2020}. The lattice distortion therefore enables the system to gain electronic energy by opening up a single-particle gap, such as in the case of a Peierls transition. While this second scenario of a hybridization-gap insulator is distinct from the first scenario of an excitonic insulator, a common feature is that they both require low-energy electronic excitations. 

Isostructural Ta$_2$NiS$_5$ serves as a valuable foil to better understand Ta$_2$NiSe$_5$ \cite{DiSalvo_1986}. Substitution of Se for S in Ta$_2$Ni(S$_x$Se$_{1-x}$)$_5$ opens/enlarges a band gap in the high-temperature normal state, which exceeds 0.25 eV in Ta$_2$NiS$_5$ \cite{Larkin_2017}. As Ta$_2$Ni(S$_x$Se$_{1-x}$)$_5$ becomes more insulating with increasing $x$, $T_\textrm{c}$ falls and signatures of an electronic transition, e.g., in electrical transport, become harder to resolve. Initially, it was thought that above $x \approx 0.55$, $T_\textrm{c}$ is fully suppressed to $<$2 K \cite{Lu_NatComm_2017}. More recent experiments using Raman spectroscopy \cite{Volkov_PRB_2021, Ye_2021} and synchrotron x-ray diffraction \cite{Chen_NatComm_2023} have reported that an orthorhombic-to-monoclinic transition persists beyond $x \approx 0.55$, even when the electronic transition is nearly undetectable, and that $T_\textrm{c}$ remains perhaps as high as 120 K at $x$ = 1 \cite{Volkov_PRB_2021, Ye_2021}. The existence of a structural transition in Ta$_2$NiS$_5$, where electronic degrees of freedom are mostly absent, points to an intrinsic lattice instability in the Ta$_2$Ni(S$_x$Se$_{1-x}$)$_5$ family as a third possible factor contributing to their phase transition. This three-instabilities picture---excitonic, hybridization-gap, and intrinsic lattice instabilities---raises renewed questions concerning their relative importance and interplay. In light of the common structural transition in Ta$_2$NiSe$_5$ and Ta$_2$NiS$_5$, it has even been questioned whether electronic instabilities are an essential part of the story \cite{Windgaetter_2021}.

Thermal transport is ideally suited to probe the dual electronic and structural transition in Ta$_2$NiSe$_5$. It is sensitive not only to electrons and holes, but also charge-neutral carriers, such as phonons and excitons. It further probes the mutual coupling and scattering of these carriers. Previously, thermal conductivity revealed an unusual anomaly in Ta$_2$NiSe$_5$ at $T_\textrm{c}$, where phonons are strongly scattered, but only in the direction perpendicular to the Ta/Ni chains \cite{Zhang_PRB_2021}. It was postulated that this anisotropic scattering could be caused by a soft transverse acoustic phonon, which is associated with the monoclinic distortion and has momentum perpendicular to the chains. Here, we extend these thermal transport experiments to track the $T_\textrm{c}$ anomaly across the entire Ta$_2$Ni(S$_x$Se$_{1-x}$)$_5$ family. Along with electrical transport and magnetic susceptibility measurements, we reveal that the soft-phonon transport anomaly is closely tied to underlying electronic instabilities, and serves as a tell-tale feature that further distinguishes Ta$_2$NiSe$_5$ from Ta$_2$NiS$_5$. Based upon our results, we present a picture for the three instabilities throughout the Ta$_2$Ni(S$_x$Se$_{1-x}$)$_5$ phase diagram. 

\textit{Experiments}---We synthesized single crystals of Ta$_2$Ni(S$_x$Se$_{1-x}$)$_5$ with $x$ = 0, 0.2, 0.5, 0.9, and 1 via chemical vapor transport \cite{Lu_NatComm_2017}. For the batches with mixed Se/S content, the results of energy-dispersive x-ray spectroscopy on representative crystals confirmed our target $x$ values \cite{SM}. Using heat capacity as a thermodynamic probe, we determined $T_c$ = 326, 303, 234, and 145 K for the $x$ = 0, 0.2, 0.5, and 0.9 samples, respectively. We could not detect any transition feature in the $x$ = 1 samples. We performed transport on rectangular-shaped crystals with the electric or heat current applied either parallel ($\parallel$; along the $a$ axis) or perpendicular ($\perp$; along the $c$ axis) to the Ta/Ni chains [Fig.~\ref{Fig1}(b)]. 

%We also measured the magnetic susceptibility of our samples.

\textit{Results}---We first survey the overall electrical and thermal transport properties of the end members of the Ta$_2$Ni(S$_x$Se$_{1-x}$)$_5$ family, i.e., Ta$_2$NiSe$_5$ and Ta$_2$NiS$_5$. Figures \ref{Fig1}(c) and \ref{Fig1}(d) present Arrhenius plots of their temperature-dependent electrical conductivities, $\sigma_\parallel(T)$ and $\sigma_\perp(T)$. Close to 400 K, $\sigma_\parallel$ of Ta$_2$NiSe$_5$ is 8$\times$10$^{4}$ $\Omega^{-1}$m$^{-1}$, which is typical of a poor metal or narrow-gap semiconductor. Its $\sigma_\perp$ is an order of magnitude smaller, 5$\times$10$^{3}$ $\Omega^{-1}$m$^{-1}$, which indicates that its electrical transport is quasi-1D along the chain direction. Ta$_2$NiS$_5$ has both smaller $\sigma_\parallel$ and $\sigma_\perp$ near 400 K, 5$\times$10$^{3}$ and 1$\times$10$^{3}$ $\Omega^{-1}$m$^{-1}$, respectively, which confirms that it is more insulating than Ta$_2$NiSe$_5$ at high temperatures. Below 400 K, the logarithms of $\sigma_\parallel$ and $\sigma_\perp$ of both compounds decrease roughly linearly with the inverse temperature $T^{-1}$, which is somewhat reminiscent of activated transport. Below 100 K, however, the decrease slows down, indicating that the intrinsic electrical conductivities are likely shorted by impurity-band contributions [blue shaded regions in Figs.~\ref{Fig1}(c) and \ref{Fig1}(d)]. For Ta$_2$NiSe$_5$, a kink is observed at $T_\textrm{c}$ = 326 K, below which $\sigma_\parallel$, and to a smaller extent $\sigma_\perp$, drop more rapidly [inset of Fig.~\ref{Fig1}(c)]. 

Figures \ref{Fig1}(e) and \ref{Fig1}(f) display the thermal conductivities $\kappa_{\parallel}(T)$ and $\kappa_{\perp}(T)$ of Ta$_2$NiSe$_5$ and Ta$_2$NiS$_5$. The thermal conductivities of Ta$_2$NiSe$_5$ are roughly 2--4 times larger than those of Ta$_2$NiS$_5$. For both compounds, $\kappa_{\parallel}$ is also roughly 2--4 times larger than $\kappa_{\perp}$. Below 400 K, the thermal conductivities increase until they reach a broad peak around 40--70 K, then decrease with further lowering of temperature. The broad peak is characteristic of phonon thermal conductivity and appears when the conductivity of phonons is no longer limited by their mean free path at higher temperatures, but by their thermal population at lower temperatures \cite{Tritt_2005}. 

\begin{figure*}
\includegraphics[width=\textwidth]{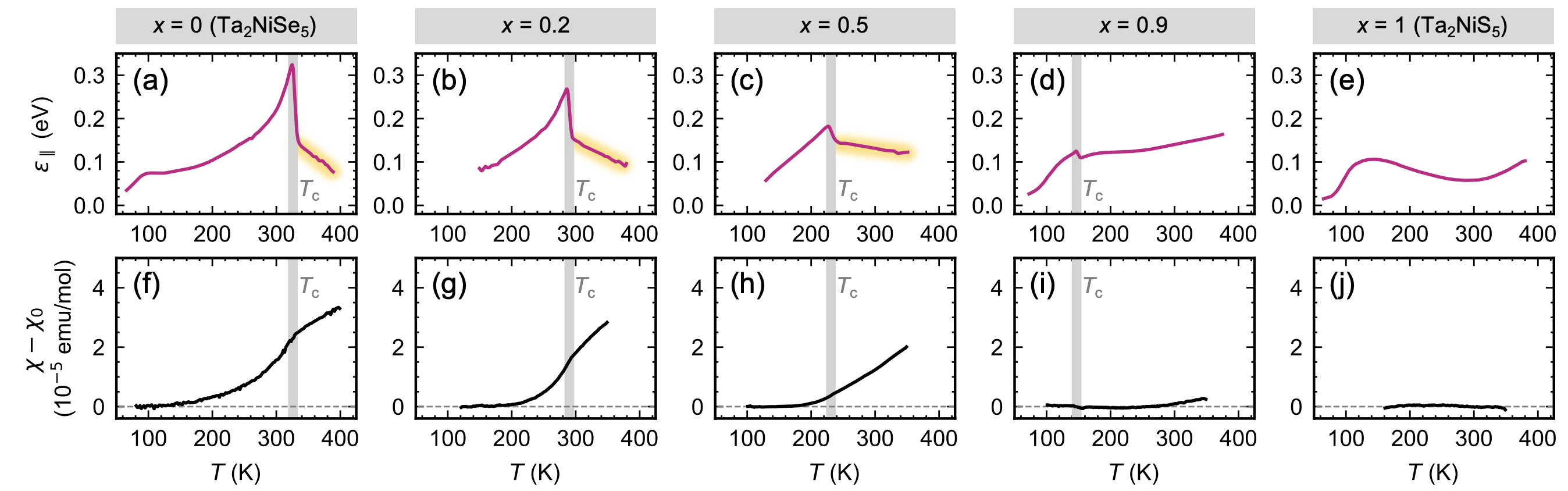}
\caption{Signatures of electronic gap opening revealed by a combination of intrachain electrical transport and magnetic susceptibility. (a)--(e) Logarithmic derivative of the intrachain electrical conductivity, $\epsilon_\parallel = k_\textrm{B} T^2 d (\ln \sigma_\parallel)/dT$, of Ta$_2$Ni(S$_x$Se$_{1-x}$)$_5$ samples with $x$ = 0, 0.2, 0.5, 0.9, and 1. The vertical gray lines mark $T_\textrm{c}$. The yellow shading in (a)--(c) marks regions of possible gap fluctuations. (f)--(j) Magnetic susceptibility $\chi - \chi_0$ vs.~$T$, where $\chi_0$ is a constant offset subtracted from the data. The data in (a) are replotted from Ref.~\cite{Zhang_PRB_2021}.}
\label{Fig2}
\end{figure*}

The measured heat current contains both electronic and phonon contributions. To estimate the electronic contribution $\kappa^{(\textrm{e})}$ to the measured $\kappa$, we apply the Wiedemann-Franz law, $\kappa^{(\textrm{e})} = L \sigma T$, where $L \approx L_0$, the Lorenz number. The phonon contribution $\kappa^{(\textrm{ph})}$ is given by the difference $\kappa - \kappa^{(\textrm{e})}$. For $\kappa_{\perp}$ of Ta$_2$NiSe$_5$ and both $\kappa_{\parallel}$ and $\kappa_{\perp}$ of Ta$_2$NiS$_5$, the estimated $\kappa^{(\textrm{e})}$ is less than one percent of $\kappa$, and $\kappa$ can be taken simply as $\kappa^{(\textrm{ph})}$. For $\kappa_{\parallel}$ of Ta$_2$NiSe$_5$, however, $\kappa^{(\textrm{e})}$ reaches around 10\% of $\kappa$ at high temperatures [compare circle symbols and dashed curves in Fig.~\ref{Fig1}(e)]. As shown in the Fig.~\ref{Fig1}(e) inset, there is a clear separation between $\kappa_{\parallel}$ and $\kappa^{(\textrm{ph})}_{\parallel}$ near $T_\textrm{c}$ = 326 K. The small dip in $\kappa_{\parallel}$ observed below $T_\textrm{c}$ is inherited from the similar dip in $\sigma_{\parallel}$ (Fig.~\ref{Fig1}(c) inset), for when we subtract $\kappa^{(\textrm{e})}_{\parallel}$ from $\kappa_{\parallel}$, the resulting $\kappa^{(\textrm{ph})}_{\parallel}$ evolves smoothly across $T_\textrm{c}$. In our later discussion, we focus on the phonon thermal conductivity and subtract $\kappa^{(\textrm{e})}$ from $\kappa$ whenever it exceeds a few percent of $\kappa$.       

We proceed to track signatures of the phase transition across the Ta$_2$Ni(S$_x$Se$_{1-x}$)$_5$ family. Figures \ref{Fig2}(a)--\ref{Fig2}(e) show the intrachain electrical conductivities of samples from our five different Se/S compositions, plotted as the logarithmic derivative $\epsilon_\parallel = k_\textrm{B} T^2 d (\ln \sigma_\parallel)/dT$, where $k_\textrm{B}$ is Boltzmann's constant. When the electrical conductivity strictly follows activated transport, i.e., $\sigma$ = $\sigma_0 \exp [E_\textrm{A}/(k_\textrm{B} T)]$, $\epsilon_\parallel$ should be a constant that is equal to the activation energy $E_\textrm{A}$. Here, we observe more complicated behavior. As seen in Fig.~\ref{Fig2}(a), $\epsilon_\parallel$ of Ta$_2$NiSe$_5$ exhibits a sudden jump at $T_c$ = 326 K, which corresponds to the opening of an electronic gap. Below $T_c$, $\epsilon_\parallel$ slowly decreases and saturates to $\sim$0.08 eV, before it shows a small hump around 100 K and decreases further due to impurity-band conduction. Using the jump in $\epsilon_\parallel$ as a signature of the electronic transition, we determine $T_\textrm{c}$ to be 290, 231, and 147 K for the $x$ = 0.2, 0.5, and 0.9 samples [Figs.~\ref{Fig2}(b)--\ref{Fig2}(d)], which closely match the $T_\textrm{c}$ values we determined from heat capacity \cite{SM}. As $x$ increases, we observe not only a decrease in $T_\textrm{c}$, but also a decrease in the size of the jump in $\epsilon_\parallel$, which implies that the electronic gap that opens below $T_\textrm{c}$ gets smaller. At $x$ = 1, we no longer observe any discontinuous feature in $\epsilon_\parallel$. It is both possible that the electronic transition is too weak to be detected, and that its signature is partially masked by the impurity-band hump around 140 K [Fig.~\ref{Fig2}(e)].             

To confirm that the jump in $\epsilon_\parallel$ at $T_\textrm{c}$ truly originates from electronic gap opening, we present complimentary magnetic susceptibility data $\chi(T)$ that also reveal signatures of electronic gap opening. The data in Figs.~\ref{Fig2}(f)--\ref{Fig2}(j) are shown after removing a Curie tail due to paramagnetic impurities. We subtract a constant offset $\chi_0$, which varies between $-2.0\times10^{-4}$ and $-1.2\times10^{-4}$ emu/mol \cite{SM}. This range of $\chi_0$ lies reasonably close to the expected orbital diamagnetism of core electrons in Ta$_2$NiSe$_5$ and Ta$_2$NiS$_5$, which we estimate to be $-2.8\times10^{-4}$ and $-1.9\times10^{-4}$ emu/mol, respectively. What remains in $\chi - \chi_0$ is predominantly the Pauli paramagnetism of quasi-1D electronic carriers. For the $x$ = 0, 0.2, and 0.5 samples, $\chi - \chi_0$ begins on the order of several $+10^{-5}$ emu/mol at high temperatures, then gradually decreases upon lowering temperature. Below $T_\textrm{c}$, a more rapid dip ensues, then $\chi - \chi_0$ appears to saturate at zero. As Pauli paramagnetism is proportional to the carrier density, it should be suppressed whenever a gap is present. We therefore ascribe the dip in $\chi - \chi_0$ seen below $T_\textrm{c}$ to the opening of an electronic gap, which again gets reduced as $x$ increases. (We note that the Knight shift of Ta$_2$NiSe$_5$, which also probes the Pauli paramagnetic spin susceptibility, shows a similar dip below $T_\textrm{c}$ \cite{Li_PRB_2018}.) For the $x$ = 0.9 and 1.0 samples, $\chi - \chi_0$ is almost negligible, which suggests that they already have a sizable band gap above $T_\textrm{c}$. There is a possible kink close to $T_\textrm{c}$ = 147 K in the $x$ = 0.9 sample, but $\chi - \chi_0$ otherwise shows little change.

\begin{figure*}
\includegraphics[width=\textwidth]{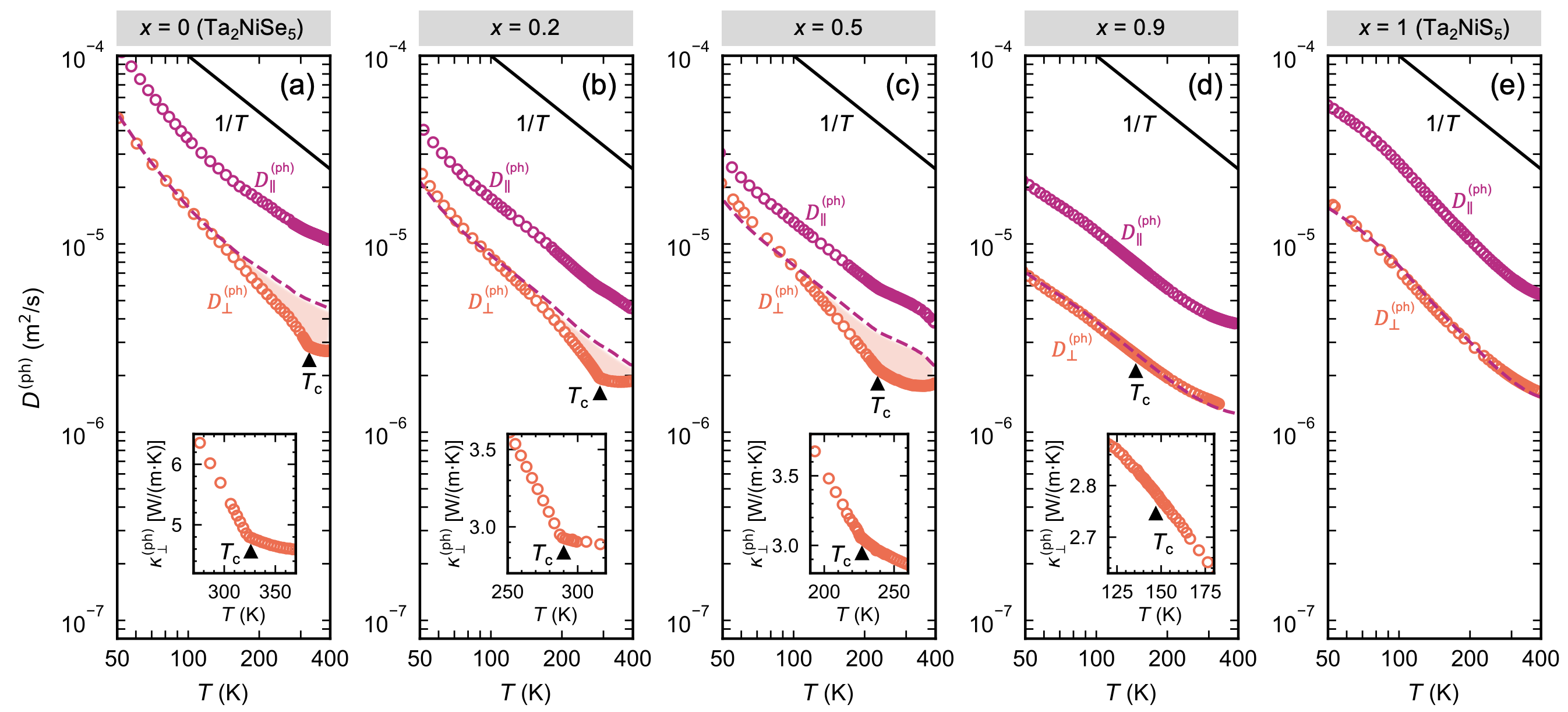}
\caption{Signature of soft-phonon anomaly revealed by interchain thermal transport. (a)--(e) Phonon diffusivity $D^{(\textrm{ph})}$ = $\kappa^{(\textrm{ph})}/C^{(\textrm{ph})}$ for Ta$_2$Ni(S$_x$Se$_{1-x}$)$_5$ samples with $x$ = 0, 0.2, 0.5, 0.9, and 1. The dashed fuchsia curves are $D_\parallel^{(\textrm{ph})}$ scaled by a prefactor to roughly match $D_\perp^{(\textrm{ph})}$, and the orange shaded regions in (a)--(c) illustrate the relative suppression of $D_\perp^{(\textrm{ph})}$. The black triangles mark $T_\textrm{c}$. The solid black lines mark the $T^{-1}$ power law expected for phonon-phonon Umklapp scattering. The insets in (a)--(d) show the $T_\textrm{c}$ anomalies (or lack thereof) in $\kappa_\perp^{(\textrm{ph})}$. The data in (a) are replotted from Ref.~\cite{Zhang_PRB_2021}.}
\label{Fig3}
\end{figure*}

We next track the transition signature in the phonon transport of Ta$_2$Ni(S$_x$Se$_{1-x}$)$_5$. Figure \ref{Fig3} plots the phonon thermal conductivities $\kappa^{(\textrm{ph})}_{\parallel}(T)$ and $\kappa^{(\textrm{ph})}_{\perp}(T)$, normalized by the phonon heat capacity $C^{(\textrm{ph})}(T)$, on double-logarithmic scales. According to the kinetic theory of phonons, $\kappa^{(\textrm{ph})}/C^{(\textrm{ph})}$ yields the diffusivity $D^{(\textrm{ph})}$ and is equal to $\frac{1}{3}v l$, where $v$ is the group velocity of phonons and $l$ is their mean free path \cite{Berman_1976}. As seen in Fig.~\ref{Fig3}, $D^{(\textrm{ph})}_{\parallel}$ for all samples decreases relatively smoothly with increasing temperature between 50 and 400 K. The temperature dependence resembles a $T^{-1}$ power law expected for $l$ at high temperatures, when phonon-phonon Umklapp scattering is dominant \cite{Berman_1976}. In contrast, $D^{(\textrm{ph})}_{\perp}$ for the $x$ = 0, 0.2, and 0.5 samples exhibits a prominent anomaly at $T_\textrm{c}$ [Fig.~\ref{Fig3}(a)--\ref{Fig3}(c)]. This anomaly originates from $\kappa^{(\textrm{ph})}_{\perp}$ (see insets of Fig.~\ref{Fig3}), not $C^{(\textrm{ph})}(T)$, as the latter evolves smoothly with temperature \cite{SM}. This anomaly is characterized by an abrupt slope change at $T_\textrm{c}$, as well as a deviation from $T^{-1}$ behavior for a temperature range over $\pm$100 K above and below $T_\textrm{c}$. When we rescale $D^{(\textrm{ph})}_{\parallel}$ on top of $D^{(\textrm{ph})}_{\perp}$ (dashed fuchsia curves), we find that $D^{(\textrm{ph})}_{\perp}$ shows an asymmetric V-shaped suppression centered at $T_\textrm{c}$ (orange shaded regions). For the $x$ = 0.9 and 1 samples, this transition anomaly vanishes. As seen in Figs.~\ref{Fig3}(d) and \ref{Fig3}(e), $D^{(\textrm{ph})}_{\parallel}$ scales on top of $D^{(\textrm{ph})}_{\perp}$ and both show approximate $T^{-1}$ dependence from 50 to 400 K. 

\textit{Discussion}---What is the microscopic origin of the V-shaped suppression in phonon thermal transport perpendicular to the Ta/Ni chains, and the significance of its disappearance at high $x$? The suppression cannot originate from the direct scattering of phonons by electronic carriers, as the latter have momenta primarily along the chains. We instead trace the origin of this suppression to a transverse acoustic phonon mode, whose momentum is perpendicular to the chains. According to x-ray scattering, this mode already softens in Ta$_2$NiSe$_5$ at 400 K, far higher than $T_\textrm{c}$ \cite{Nakano_PRB_2018, Kang_arXiv_2025}. As phonon diffusivity is the product of $v$ and $l$, the softening of this mode could suppress $D^{(\textrm{ph})}_{\perp}$ in two ways, via its own reduced group velocity, and/or via its enhanced scattering of the other phonons. Since Ta$_2$NiSe$_5$ has 48 phonon branches, many of which are dispersive \cite{Nakano_PRB_2018, Subedi_2020}, we do not expect the reduced velocity of one soft phonon to greatly influence $D^{(\textrm{ph})}_{\perp}$. The dominant mechanism behind the suppression of $D^{(\textrm{ph})}_{\perp}$ should be the increased scattering of phonons by this soft mode, resulting in an overall decrease of the mean free path. In real space, this soft phonon corresponds to shearing fluctuations of the chains that precede the monoclinic distortion. In this fluctuating state, the individual chains remain largely unchanged, but are displaced in the $a$ direction relative to their neighbors. Such fluctuations would strongly scatter phonons propagating perpendicular to the chains, while minimally perturbing phonons propagating parallel to the chains, consistent with our observations in Fig.~\ref{Fig3}. We note that a similar V-shaped suppression in phonon thermal conductivity has been observed in the order-disorder-type ferroelectrics KH$_2$PO$_4$ and NH$_4$Cl, where fluctuating local dipoles give rise to a soft mode \cite{Bausch_PLA_1969, Suemune_JPSJ_1967}.

Although the transition anomaly in $\kappa^{(\textrm{ph})}_{\perp}$ is related to a soft transverse acoustic phonon, its dependence on the Se/S composition, which tunes normal-state band overlap/gap, indicates that the soft phonon is electronically tuned.  The fact that the soft-phonon transport anomaly becomes weaker as signatures of electronic gap opening in $\epsilon_\parallel$ and $\chi$ also get weaker establishes a close relationship between the soft phonon and underlying electronic instabilities. This connection is further affirmed by previous reports of electronic gap fluctuations in Ta$_2$NiSe$_5$ above $T_\textrm{c}$, in the same regime where the soft phonon is observed. According to optical and photoemission spectroscopy, the electronic gap does not entirely close at $T_\textrm{c}$; instead, the spectral weight near the Fermi energy recovers slowly above $T_\textrm{c}$, indicating a fluctuating gap \cite{Lu_NatComm_2017, Chen_PRR_2023}. A fluctuating gap could possibly explain the slow rise in $\epsilon_\parallel$ as the temperature decreases toward $T_\textrm{c}$, just prior to the sudden jump, as seen in the $x$ = 0, 0.2, and 0.5 samples [yellow shading in Figs.~\ref{Fig2}(a)--\ref{Fig2}(c)]. We infer that the soft phonon with shearing lattice fluctuations is accompanied by and coupled to electronic fluctuations.

\begin{figure}
\includegraphics[width=\columnwidth]{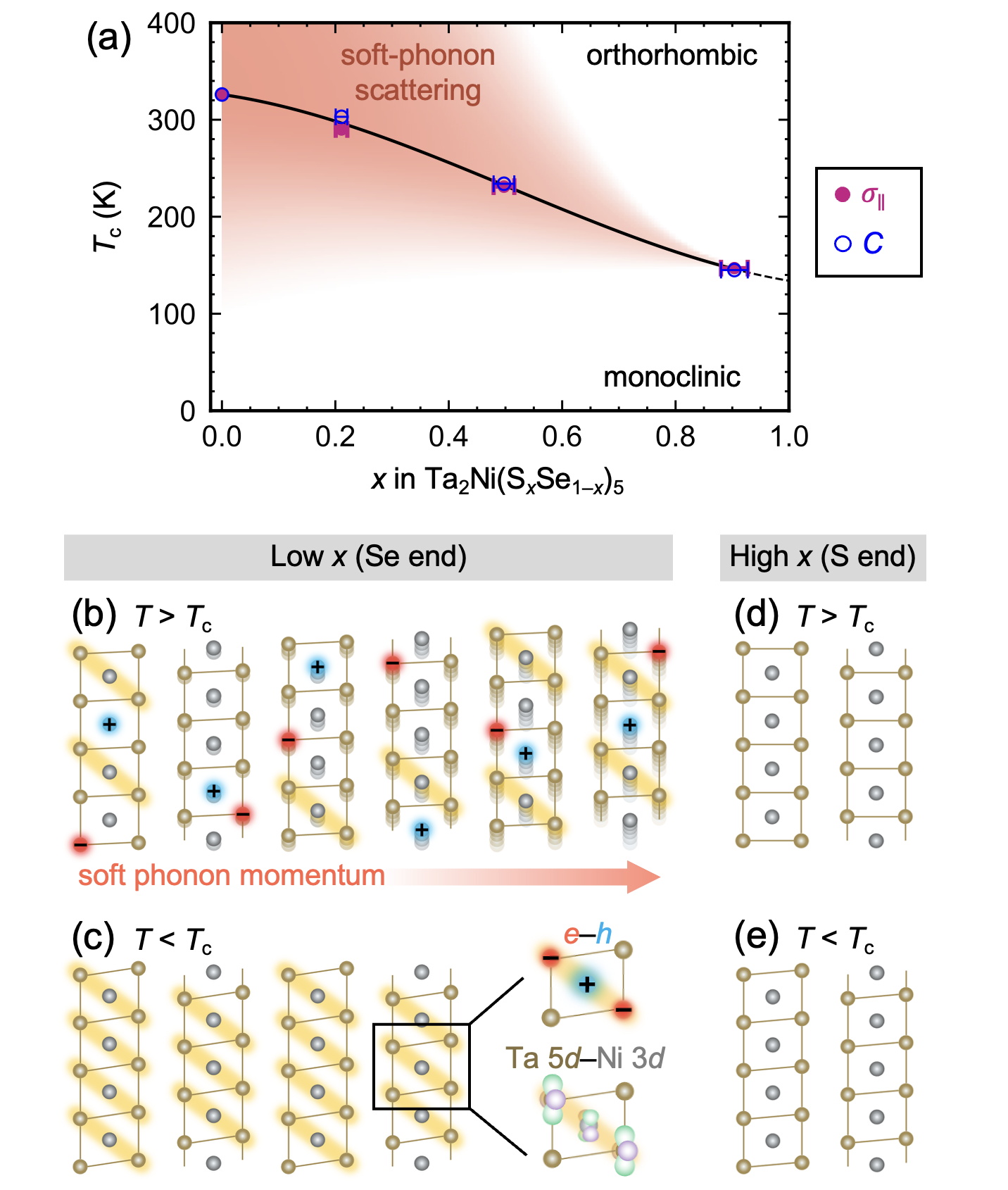}
\caption{(a) Phase diagram of Ta$_2$Ni(S$_x$Se$_{1-x}$)$_5$. The filled fuchsia and open blue circles are $T_\textrm{c}$ values derived from the intrachain electrical conductivity and heat capacity, respectively. The orange shading marks the region of enhanced phonon-phonon scattering in the interchain direction. (b), (c) Schematic of Ta/Ni chains above and below $T_\textrm{c}$, respectively, near the Se end. The faint yellow bars represent electronic interactions between the Ta and Ni sites, which could be the pairing of electrons and holes, or the hybridization of Ta 5$d$ and Ni 3$d$ orbitals [illustrated in (c)]. The monoclinic distortion is exaggerated for clarity. (d), (e) Schematic of Ta/Ni chains above and below $T_\textrm{c}$, respectively, near the S end.}
\label{Fig4}
\end{figure}

Figure \ref{Fig4} summarizes our updated picture of the Ta$_2$Ni(S$_x$Se$_{1-x}$)$_5$ family. We plot $T_\textrm{c}$ as a function of $x$ and highlight the region of soft-phonon scattering seen in thermal transport [Fig.~\ref{Fig4}(a)]. At $x$ = 1, the phase transition is apparently too weak to be detected by any of our probes. If we loosely extrapolate our $T_\textrm{c}$ line to $x$ = 1 (dotted curve), we would estimate $T_\textrm{c}$ $\approx$ 130 K in Ta$_2$NiS$_5$. This value is close to the $T_\textrm{c}$ of 120 K determined by polarized Raman spectroscopy, which is extremely sensitive to the difference between orthorhombic and monoclinic lattice symmetries \cite{Volkov_PRB_2021, Ye_2021}.      

% We concurrently plot the size of the jump in $\epsilon_\parallel$, which is a measure of the strength of the underlying electronic instabilities. 

Our results distinguish the nature of the phase transition at the Se and S ends of the Ta$_2$Ni(S$_x$Se$_{1-x}$)$_5$ family, as schematically illustrated in Figs.~\ref{Fig4}(b)--\ref{Fig4}(e). At the S end, we begin with a sizable band gap in the normal state with electronic degrees of freedom minimized. There is apparently an intrinsic lattice instability giving rise to a weak orthorhombic-to-monoclinic distortion at $T_\textrm{c} \approx 130$ K, one that is invisible to most experiments. Moving toward the Se end, as the normal-state band gap closes, additional electronic instabilities emerge. These electronic instabilities not only boost the structural transition temperature from $\sim$130 to 326 K, but also intensify the experimental signatures of $T_\textrm{c}$ in both electronic and structural channels. The phase transition of the $x \leq 0.5$ samples is characterized not only by electronic gap opening and a monoclinic distortion at $T_\textrm{c}$, but also by a soft-phonon transport anomaly with electronic fluctuations over a much wider temperature range. These latter features are notably absent in the $x$ = 0.9 and 1 samples.

What is the nature of the electronic fluctuations that accompany the soft-phonon transport anomaly? They could be preformed excitons coupled to the lattice, which give rise a fluctuating many-body gap, as suggested by a photoemission report \cite{Fukutani_NatPhys_2021}. They could equally well be fluctuations in the single-particle hybridization gap between the Ta 5$d$ and Ni 3$d$ bands coupled to the lattice, similar to charge-density-wave fluctuations \cite{Lee_PRL_1973, Johnston_1985}. Both scenarios have garnered experimental support \cite{Wakisaka_2009, Watson_PRR_2020, Kim_NatComm_2021, Fukutani_NatPhys_2021, Chen_NatComm_2023, Golez_2022, Katsumi_2023, Baldini_PNAS_2023, Shi_PRL_2025, Rosenberg_PRL_2026}, and neither should be fully excluded from the discussion. We would like to propose that in fact, all three instabilities---excitonic, hybridization-gap, and intrinsic lattice instabilities---play an indispensable role in Ta$_2$NiSe$_5$, and should be simultaneously considered in any quantitative modeling. We give the following justification: From our Ta$_2$Ni(S$_x$Se$_{1-x}$)$_5$ phase diagram, we estimate the temperature scale of the intrinsic lattice instability to be $\sim$130 K, which is sizable, but falls short of the maximum $T_\textrm{c}$ of 326 K. From analyses of electronic Raman scattering \cite{Volkov_2021, Volkov_PRB_2021, Ye_2021, Zhang_arXiv_2025}, a bare excitonic instability with temperature scale of $\sim$137 K has been suggested, which again is a sizable, but falls short of 326 K. From density functional theory, a single-particle hybridization gap of 0.12 eV has been calculated \cite{Watson_PRR_2020}, which is sizable, but nevertheless falls short of the experimental charge gap, 0.16--0.30 eV \cite{Lu_NatComm_2017, Lee_PRB_2019, He_2021}. Clearly, each of the three instabilities is justified, yet none is sufficient alone in isolation. It is their collaboration that is ultimately responsible for the high $T_\textrm{c}$ of 326 K in Ta$_2$NiSe$_5$.

\textit{Acknowledgements}---We would like to thank C.~Busch, M.~Dueller, and K.~Fink for technical support and Y.~Matsumoto, T.~T.~M.~Palstra, and P.~Reiss for discussions. This work was partly supported by the Deutsche Forschungsgemeinschaft (DFG, German Research Foundation) -- TRR 360 -- 492547816.

%\bibliography{References_Ta2NiSe5}


\end{document}